\documentclass[
    prapplied,
    preprint,
    reprint,
    superscriptaddress,
    twocolumn
    ]{revtex4-2}

\usepackage{amsmath,amssymb}
\usepackage{graphicx}
\graphicspath{{figures/}}
\usepackage{dcolumn}
\usepackage{bm}
\usepackage{ulem}
\usepackage{soul}
\usepackage[utf8]{inputenc}
\usepackage[T1]{fontenc}
\usepackage{mathptmx}
\usepackage{etoolbox}
\usepackage{siunitx}
\usepackage{lineno}
\DeclareSIUnit{\belmilliwatt}{Bm}
\DeclareSIUnit{\dBm}{\deci\belmilliwatt}
\DeclareUnicodeCharacter{0300}{}
\DeclareUnicodeCharacter{0301}{}
\DeclareUnicodeCharacter{0327}{}

\usepackage{hyperref}
\hypersetup{
    colorlinks=true,
    linkcolor=blue,
    filecolor=magenta, 
    citecolor = blue,
    urlcolor=blue,
    pdfpagemode=FullScreen,
    }

\usepackage{color,soul}

\begin{document}

\title{Active control of THz plasmon propagation in a one-dimensional electronic waveguide}

\author{Thomas Vasselon}
\affiliation{Université Grenoble Alpes, CNRS, Grenoble INP, Institut Néel, Grenoble, 38000, France}

\author{Uzer Ahmad}
\affiliation{Université Grenoble Alpes, CNRS, Grenoble INP, Institut Néel, Grenoble, 38000, France}

\author{Clément Geffroy}
\affiliation{Université Grenoble Alpes, CNRS, Grenoble INP, Institut Néel, Grenoble, 38000, France}

\author{Lucas Mazzella}
\affiliation{Université Grenoble Alpes, CNRS, Grenoble INP, Institut Néel, Grenoble, 38000, France}

\author{Matteo Aluffi}
\affiliation{Université Grenoble Alpes, CNRS, Grenoble INP, Institut Néel, Grenoble, 38000, France}

\author{Seddik Ouacel}
\affiliation{Université Grenoble Alpes, CNRS, Grenoble INP, Institut Néel, Grenoble, 38000, France}

\author{Jashwanth Shaju}
\affiliation{Université Grenoble Alpes, CNRS, Grenoble INP, Institut Néel, Grenoble, 38000, France}

\author{Kevin Bredillet}
\affiliation{Universit\'e Grenoble Alpes, Universit\'e Savoie Mont-Blanc, CNRS, Grenoble INP, CROMA, Grenoble, 38000, France}

\author{Nathan Roussel}
\affiliation{Universit\'e Grenoble Alpes, Universit\'e Savoie Mont-Blanc, CNRS, Grenoble INP, CROMA, Grenoble, 38000, France}

\author{Jean-François Roux}
\affiliation{Universit\'e Grenoble Alpes, Universit\'e Savoie Mont-Blanc, CNRS, Grenoble INP, CROMA, Grenoble, 38000, France}

\author{Pierre-Baptiste Vigneron}
\affiliation{Universit\'e Grenoble Alpes, Universit\'e Savoie Mont-Blanc, CNRS, Grenoble INP, CROMA, Grenoble, 38000, France}

\author{Arne Ludwig}
\affiliation{Lehrstuhl für Angewandte Festkörperphysik, Ruhr-Universität Bochum, Bochum, Germany}

\author{Andreas D. Wieck}
\affiliation{Lehrstuhl für Angewandte Festkörperphysik, Ruhr-Universität Bochum, Bochum, Germany}

\author{Matias Urdampilleta}
\affiliation{Université Grenoble Alpes, CNRS, Grenoble INP, Institut Néel, Grenoble, 38000, France}

\author{Hermann Sellier}
\affiliation{Université Grenoble Alpes, CNRS, Grenoble INP, Institut Néel, Grenoble, 38000, France}

\author{Giorgos Georgiou}
\affiliation{James Watt School of Engineering, Electronics and Nanoscale Engineering, University of Glasgow, Glasgow, G12 8QQ, United Kingdom}

\author{Christopher B\"auerle}
\thanks{Corresponding Author: christopher.bauerle@neel.cnrs.fr}
\affiliation{Université Grenoble Alpes, CNRS, Grenoble INP, Institut Néel, Grenoble, 38000, France}

\begin{abstract}
Quantum nanoelectronics is pushing towards ever higher operating frequencies in order to realise quantum technologies capable of processing information at unprecedented speeds. 
A particularly promising direction is the development of flying electron qubits, which offer the prospect of quantum operations on picosecond timescales. 
Achieving in-flight quantum control in this regime would establish a fundamentally new paradigm for studying quantum entanglement and enable a novel form of quantum information processing based on propagating electronic wavepackets.
Here we report a first step towards this goal by injecting ultrashort electron wavepackets into an engineered quantum nanoelectronic device. 
We demonstrate active control over the propagation speed of an electron wavepacket with a temporal duration as short as \SI{4}{\pico\second} in a quasi-one-dimensional electron waveguide with a length ranging from \SIrange{10}{40}{\micro\meter}.
This advance provides a key building block for ultrafast quantum operations using flying electrons. 
Beyond its technological implications, our approach offers a platform for exploring the intrinsic dynamical processes that govern quantum transport and coherence in nanoscale electronic systems. 

\end{abstract}

\maketitle

\section*{Introduction}

\noindent The relentless drive towards miniaturisation and higher operating speeds in microelectronic devices is pushing conventional semiconductor technologies to their limits. 
From kilohertz switching electronics to gigahertz processors and \SI{}{\milli\meter}-wave communications, successive increases in operating frequency have enabled transformative technological advances.
As device dimensions approach nanometre scales and signal periods enter the picosecond and sub-picosecond regime, electronic transport becomes strongly coupled to internal dynamics and many-body interactions. 
In this regime, frequency is no longer merely a performance metric; it becomes a probe and a control parameter of nanoscale devices, revealing otherwise inaccessible excitations and interactions \cite{Gaury2014, Kotilahti2021, Saha2025, Kloss2025a}.

High-frequency operation is therefore emerging as a route towards smaller, faster, and more coherent quantum devices. 
Recent progress includes ultra-low-power superconducting logic based on picosecond pulses generated by Josephson junctions \cite{Likharev1991, Liu2023},  ultrafast readout of spintronic memories enabled by picosecond electrical pulses \cite{Kampfrath2013, Jhuria2020}, and semiconductor transistors operating beyond the drift–diffusion limit, opening perspectives for ultrafast switching and high-frequency amplification \cite{Dyakonov1993, Veksler2006,BoubangaTombet2020}.
Despite these advances, the microscopic dynamics governing charge transport at terahertz (\SI{}{\tera\hertz}) frequencies and picosecond time scales remain only partially understood.

In contrast to low-frequency transport driven by quasi-static biasing, ultrashort electrical pulses \cite{Roussely2018, Roussel2021, Aluffi2023, Ouacel2025, Takada2025, Bartolomei2025, Souquet2025} excite collective plasmon-like modes, which are coherent collective charge-density oscillations.
Such excitations have recently attracted considerable interest in two-dimensional electron systems, where magneto- and acoustic-plasmon transport have been investigated \cite{Shaner2004, Kumada2013, Wu2016, Yoshioka2024}. In parallel, cavity electrodynamics experiments have demonstrated coherent coupling to confined THz plasmonic modes in van der Waals heterostructures \cite{Kipp2025}, opening new opportunities for controlling collective electronic excitations at ultrafast timescales.

The ultrafast plasmon dynamics open a pathway towards flying electron qubits, enabling information processing at speeds far beyond those achievable in conventional solid-state architectures. 
Consequently, plasmons are gaining increasing interest as a platform for electron quantum optics, where coherent charge excitations act as flying quantum states \cite{ Bocquillon2013, Dubois2013, Roussel2017, Edlbauer2022, Assouline2023}, in direct analogy to photons in photonic quantum technologies.

Realising this vision using one-dimensional quantum devices, such as electron waveguides, phase modulators and splitters, presents significant technological challenges mostly due to the lack of sources that can generate ultrashort electrical pulses with picosecond durations.  
Furthermore, the lateral dimensions of these building blocks become critical at such high frequencies, as the plasmon wavelength approaches the device dimensions. 
Although plasmon transport in this regime remains relatively unexplored, the ability to scale these structures to sub-wavelength dimensions could enable highly compact device architectures.

In this work, we demonstrate the first quantum waveguide device supporting the propagation of plasmon wavepackets, integrated with \SI{}{\tera\hertz} optoelectronic sources capable of delivering voltage pulses as short as \SI{4}{\pico\second}.
We further show that key properties of these plasmonic excitations, most notably their propagation velocity, can be actively tuned by electrically controlling the conductance of the waveguide \cite{Roussely2018, Takada2025}. 
The wavepacket propagates over a distance of \SI{100}{\micro\meter}, demonstrating the generation of electronic wavepackets whose spatial extent is smaller than the device itself.

\section*{THz quantum nanoelectronic device}

\begin{figure*}[t]
    \centering
    \includegraphics[scale = 1]{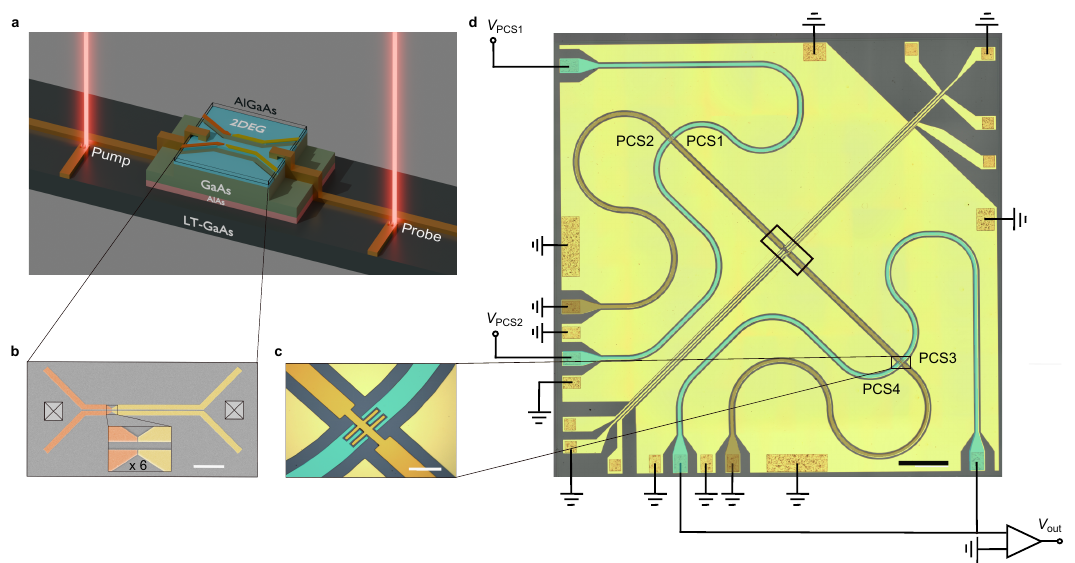}
    \caption{
    \textbf{Device architecture and chip layout}
    \textbf{a.} Schematic of the device, featuring integrated \SI{}{\tera\hertz} opto-electronics and a coplanar waveguide alongside  a quantum device fabricated on an AlGaAs/GaAs heterostructure. 
    The heterostructure was etched to expose the low-temperature-grown GaAs (LT-GaAs), where two electrodes form the photoconductive switches (PCS), labelled "pump" and "probe". 
    Both electrodes are illuminated by a \SI{780}{\nano\meter}, \SI{100}{\femto\second} laser.
    \textbf{b.} Scanning electron microscopy image of the quantum device. 
    It comprises a 1D electronic waveguide divided into two segments: a \SI{10}{\micro\meter} section (orange) and a \SI{30}{\micro\meter} section (yellow). 
    The distance between the upper and lower electrodes is \SI{500}{\nano\meter} as highlighted in the inset.
    The scale bar represents a length of \SI{10}{\micro\meter}.
    \textbf{c.} Optical microscope image of a pair of face-to-face PCS devices. 
    The PCS regions are highlighted in blue, while the central brown line represents the coplanar waveguide (CPW), which carries the picosecond electrical pulses. 
    The scale bar indicates a length of \SI{50}{\micro\meter}.
    \textbf{d.} Optical microscope image of the entire device, where the \SI{}{\tera\hertz} electronics are integrated alongside the quantum device located in the central squared area. 
    Two sets of PCS devices are used, highlighted in blue. 
    PCS1 and PCS2 serve as the pump switches, and when biased with a DC voltage they generate picosecond-duration electrical pulses.
    PCS3 and PCS4 act as the probe switches 
    and are connected to a transimpedance amplifier to measure the transmitted time-resolved signal using a pump-and-probe technique.
    The scale bar indicates a length of \SI{500}{\micro\meter}.
    }
    \label{fig1}
\end{figure*}

\noindent The device, depicted in Fig. \ref{fig1}, is designed to explore and control the propagation of \SI{}{\tera\hertz} plasmonic excitations within a one-dimensional electronic waveguide. 
As shown schematically in Fig. \ref{fig1}a, the complete device consists of two distinct parts.
The central part comprises a GaAs/AlGaAs heterostructure hosting a two-dimensional electron gas (2DEG), on which the quantum circuit is patterned. 
This circuit features a combination of \SI{10}{\micro\meter} and \SI{30}{\micro\meter}-long 
electronic waveguides, that allow for active control of the waveguide conductance between 0 and 10 conductance quanta ($G_0$ = ${2e^2}/{h}$).
At a depth of \SI{610}{\nano\meter} lies the low-temperature-grown GaAs (LT-GaAs), on which the \SI{}{\tera\hertz} electronic circuits are integrated. 
This part comprises two \SI{50}{\ohm} impedance-matched coplanar waveguides (CPW) (highlighted in brown in Fig. \ref{fig1}d), connected to the 2DEG via Ohmic contacts (grey crossed boxes in Fig. \ref{fig1}b). 
In addition, four photoconductive switches (PCS) (highlighted in blue in Fig. \ref{fig1}d) are integrated into the CPWs to generate and detect picosecond-long voltage pulses \cite{Auston1975, Smith1989}. 
The on-chip \SI{}{\tera\hertz} sources are designed with a face-to-face PCS geometry, as this configuration enables the selective injection of \SI{}{\tera\hertz} voltage pulses into the CPW's coplanar or slotline transverse electromagnetic (TEM) modes \cite{Zamdmer1999}.
By controlling the bias voltage on either side of the generating PCS, we can select the excitation of the desired coplanar mode, through symmetric biasing, and sufficiently suppress the excitation of the slotline mode (see section III in Supplementary Material).
This in turn facilitates a more efficient excitation of THz plasmons within our 2DEG system \cite{Wu2015}.

For time-resolved measurements, the laser beam is divided into two paths. 
One path, the "pump", excites the biased photoconductive switches PCS1 and PCS2.
This generates a \SI{4}{\pico\second}-short voltage pulse that propagates along the central CPW line, and injected into the 2DEG. 
The second path, the "probe", is delayed by a mechanical delay line, and focused onto PCS3 and/or PCS4 for sampling.
The resulting current is detected using a transimpedance amplifier and lock-in techniques, with the probe beam being mechanically modulated at a frequency of \SI{329}{\hertz}.
All experiments were conducted at \SI{4}{\kelvin} if not specified otherwise, with a typical optical power of approximately \SI{2}{\milli\watt} and a typical PCS bias of \SI{+12}{\volt}. \\

\begin{figure*}[t]
    \centering
    \includegraphics[scale = 1.0]{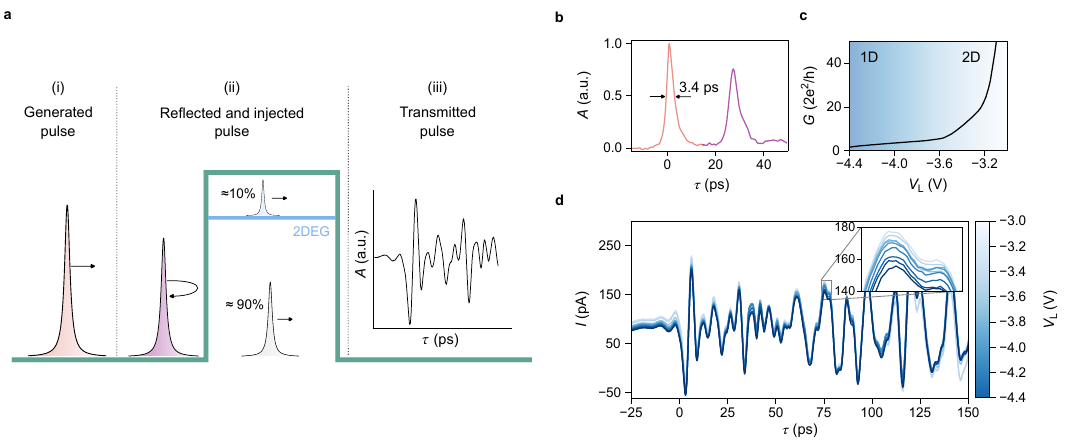}
    \caption{
    \textbf{THz voltage pulse generation and plasmon wavepacket excitation}
    \textbf{a.} Schematic illustrating the propagation of the THz voltage pulse.
    (i) Generated pulse at the photoconductive switch.
    (ii) Proportion of the generated pulse reflected at the ohmic contact, along with the injected pulse into the 2DEG and the portion propagating through the substrate. 
    (iii) Transmitted signals after the quantum device, as measured at the second set of photoconductive switches.
    \textbf{b.} Time-resolved measurement of the generated pulse (orange) just after the photoconductive switch and reflected pulse (purple) at the injection ohmic contact measured at \SI{300}{\kelvin}. 
    The generated pulse is \SI{3.4}{\pico\second} short, with a reflection accounting for \SI{75}{\percent} of the signal. 
    The measurement was performed on the coplanar waveguide (CPW) between the pump photoconductive switch (PCS) and the quantum device via sliding contact technique \cite{Grischkowsky2000}. 
    \textbf{c.} Quantum transport characterisation of the \SI{30}{\micro\meter}-long electronic waveguide. The plot shows the conductance as a function of the negative voltage applied to the electrostatic gates.
    \textbf{d.} Time-resolved measurement of the transmitted THz signal after propagation through the quantum device, detected by the probe PCS. 
    Each curve corresponds to measurements taken at different voltages applied to the long electronic waveguide. 
    The inset highlights subtle changes in the transmitted signal.
    }
    \label{fig2}
\end{figure*}

Prior to injecting the picosecond pulses into the 2DEG, we characterise the temporal quality of the pulses.
This is performed on a free-space optical THz setup at \SI{300}{\kelvin} using a sliding contact technique \cite{Grischkowsky2000}, which allowed us to measure  the temporal profile of the picosecond pulses at various points along the CPW.
This method allows the simultaneous measurement of both the pulse injected into the 2DEG and the pulse reflected from the 2DEG (for details, see Supplementary Material).
As shown in the schematic of Fig. \ref{fig2}a, at room temperature, a significant portion of the pulse is reflected at the interface between the CPW and the Ohmic contact, a result of the large impedance mismatch due to the high 2DEG resistance.
Although at cryogenic temperatures, the 2DEG resistance is orders of magnitude lower, and \SI{}{\tera\hertz} pulse transmission into the 2DEG is more efficient, through this measurement we can extract valuable quantitative information about the injected pulse duration, which it is expected to remain the same as the device is cooled at \SI{4}{\kelvin}.
The temporal duration of both generated and reflected pulses are shown in  Fig. \ref{fig2}b, with orange and purple colours respectively. 
Room-temperature characterisation of the \SI{}{\tera\hertz} pulses (see Supplementary Fig. S9) shows a slight reduction in pulse amplitude due to attenuation and a slight temporal broadening caused by the dispersion of the coplanar waveguide mode over the \SI{1.6}{\milli\meter} propagation distance from the generating PCS. As a result, the pulse duration increases from \SI{3.4}{\pico\second} at the generating PCS to \SI{4.3}{\pico\second} at the quantum device.

At cryogenic temperatures, the injected \SI{4.3}{\pico\second} short voltage pulse generates a \SI{}{\tera\hertz} plasmonic excitation inside the 2DEG that propagates through the quantum device and the time trace of the transmitted pulse is reconstructed in a time-resolved manner at PCS 3 and/or 4.
Prior to the transport measurements at \SI{}{\tera\hertz} frequencies, we proceed with device characterisation measurements at a temperature of \SI{4}{\kelvin}. 
Fig. \ref{fig2}c illustrates the conductance of the transition between 2D-to-1D, as a function of the applied negative gate voltage on the long electronic waveguide (highlighted in yellow in Fig. \ref{fig1}). 
Below a voltage of \SI{-3.6}{\volt} we observe the formation of a quasi-one-dimensional electronic waveguide. 
Similar transport measurements have been done for the short electronic waveguide (see side panel of Fig. \ref{fig3} b).
Finally, we study the transmitted \SI{}{\tera\hertz} pulse propagating through the 2DEG for different conductance values of the electronic waveguides by tuning the electrostatic gate voltages.
A typical result is shown in Fig. \ref{fig2}d. 
While a background signal is present due to \SI{}{\tera\hertz} propagation outside the 2DEG, we observe clear, reproducible variations in the transmitted current that are sensitive to the applied gate voltage (see inset).
This indicates that a portion of the \SI{}{\tera\hertz} pulse is indeed transmitted through the 2DEG and can be controlled via the electrostatic gates. 
The background contribution likely arises from direct \SI{}{\tera\hertz} transmission through the semiconductor substrate beneath the 2DEG, due to the monolithic integration of the high-mobility heterostructure with the photoactive LT-GaAs layer (see Methods).
Another explanation for the observed background is a possible capacitive coupling between the CPW transmission lines on either side of the Ohmic contacts.
This capacitive coupling would appear as an effect similar to that of a high-pass filter, with a characteristic signature of a long time-resolved signal.

To isolate the effect of \SI{}{\tera\hertz} propagation through the 2DEG, we take the derivative of the time-domain signal with respect to the voltage applied on the side gates (or equivalently, the conductance) \cite{Wu2015}. 
This procedure effectively filters out the undesired background signal and allows us to directly analyse plasmonic excitations in the one-dimensional electronic waveguide.

\section*{THz plasmon propagation within an electronic waveguide}

\begin{figure*}[t]
    \centering
    \includegraphics[scale = 1]{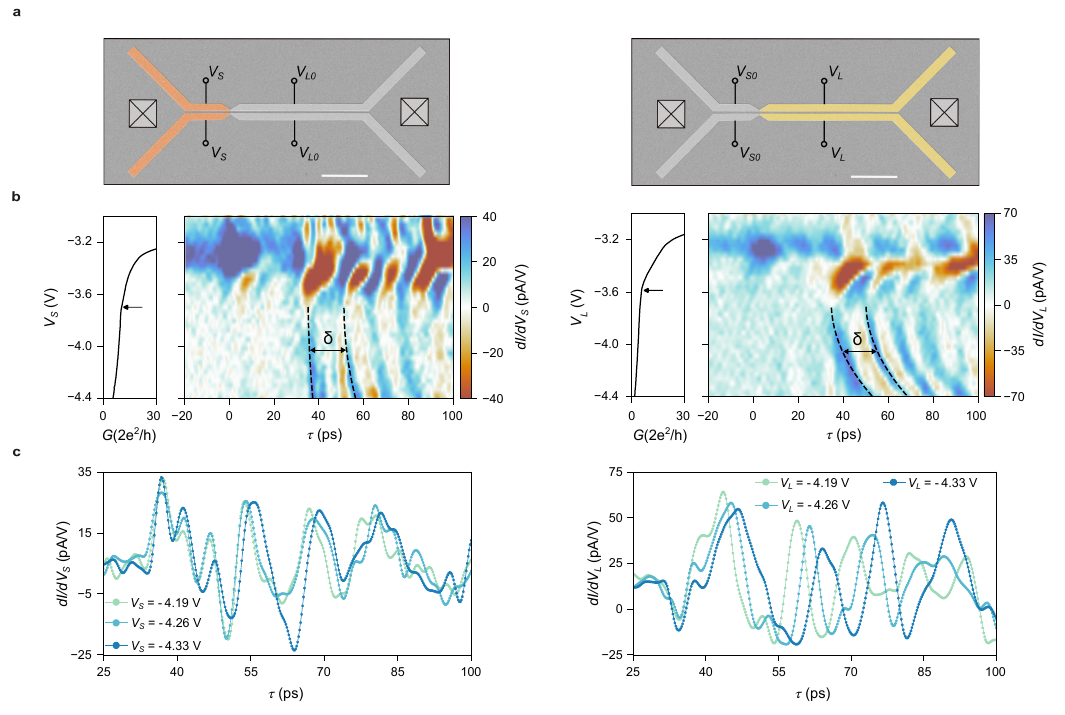}
    \caption{
    \textbf{THz Plasmonic excitations of the quantum device}
    \textbf{a.} Scanning electron microscopy image of the quantum device, with the \SI{10}{\micro\meter} (coloured in orange) and \SI{30}{\micro\meter} (coloured in yellow) long electronic waveguides. 
    Voltages $V_S$ and $V_L$ are swept to tune the conductance in a part of the wire. 
    Uncoloured gate are polarized to form a electronic waveguide at fixed conductance.
    The crossed square boxes represent the ohmic contacts, enabling electrical connection between the coplanar waveguide and the two-dimensional electron gas.
    The scale bar represents \SI{10}{\micro\meter}
    \textbf{b.} Two-dimensional map of the time-resolved transmitted signal, derived with respect to the voltage $V_S$ on the short waveguide (and $V_L$ on the long waveguide).
    The corresponding conductance of varying region is placed alongside it. 
    \textbf{c.} Horizontal cuts of the 2D map shown in \textbf{b}.
    }
    \label{fig3}
\end{figure*}

\noindent We now demonstrate active manipulation of the plasmonic wavepacket within the electronic waveguide by recording the time-resolved transmitted \SI{}{\tera\hertz} signal while tuning the wave-guide conductance. 
A clear modulation of the transmitted amplitude is observed as a function of conductance (Fig. \ref{fig2}d, inset), evidencing controlled interaction between the plasmonic excitation and the electronic confinement potential. 
As shown in Fig. \ref{fig3}a, for these measurements, we modify the conductance of the highlighted part of the electronic waveguide by sweeping either the gate voltage $V_{L}$ for the long wire, or $V_{S}$ for the short wire, while keeping the other one in the 1D regime, at fixed conductance set by $V_{L0}$ or $V_{S0}$ respectively.
The zero of the time-delay axis is defined by the first current peak, which originates from the THz pulse propagating through the substrate.

To isolate the contribution from plasmons propagating in the 2DEG, we compute the derivative of the transmitted current with respect to the applied gate voltage $V_S$ and $V_L$, as shown in Fig. \ref{fig3}a. 
Entering the 2D–1D crossover regime, the maps exhibit a clear separation of features (Fig. \ref{fig3}b), reflecting the onset of one-dimensional confinement.

Clear fringes emerge when the plasmonic excitations are confined to one dimension and importantly we also observe that as the confinement is increased, the fringes are detected at increasingly longer times.
This indicates a decrease in the plasmon propagation velocity. 
Such behaviour is consistent with earlier studies of plasmonic excitations in the \SI{}{\giga\hertz} regime \cite{Roussely2018, Takada2025}.

By performing measurements in two complementary configurations -- (i) fixing the conductance of the short waveguide while varying the conductance of the long waveguide, and (ii) fixing the conductance of the long waveguide while varying that of the short waveguide -- we are able to determine the propagation velocity $v_{\mathrm{1D}}$ over the full conductance range from $2G_0$ to $11G_0$, where $G_0$ is the quantum conductance.

To extract the propagation velocity of the plasmonic excitation in our quasi-1D electronic waveguide, we analyse the time delay $\tau$ of the signal as the conductance of the waveguide is varied. 
For this purpose, we introduce $\delta$, which is the relative temporal separation between successive interference fringes, as shown in Fig. \ref{fig3} and Supplementary Fig. S16.

Our analysis shows that the observed fringes originate from multiple reflections within the short waveguide, with additional reflections occurring at the ohmic contact where the metallic leads are connected to the 2DEG. Each round trip increases the effective propagation distance, leading to an increase of the temporal separation $\delta$ (see Supplementary Information for details).

The resulting conductance dependence of the plasmon velocity $v_{1D}$  is shown in Fig. 4. At high conductance ($G = 10\,G_0$), the velocity reaches approximately
\SI{1.7e6}{\meter\per\second}. 
As the number of conducting channels is reduced, the velocity decreases continuously, reaching approximately 
\SI{0.7e6}{\meter\per\second} at $G = 2\,G_0$. 
The observed conductance dependence follows a square-root scaling, in agreement with previous experiments in the GHz regime \cite{Roussely2018, Takada2025, Kukushkin2005}, and demonstrates that this behaviour persists into the THz frequency range.

In our device, the electronic waveguide has a width of a few hundred nanometres (see Methods), whereas the wavelength of the THz plasmonic excitations ranges from \SIrange{1}{10}{\micro\meter}.
Consequently, the electric field is strongly confined along the plane transverse to the propagation direction, resulting in the formation of effectively one-dimensional plasmonic modes.

\subsection*{Propagation velocity of a THz plasmonic excitation}

In one dimension the dispersion relation is linear \cite{Kukushkin2005} and to describe our system, we combine a transmission-line model for gated 2DEG \cite{Aizin2012, Dyer2012, Yoon2014}, with the dispersion relation of one-dimensional plasmons \cite{Kukushkin2005}. 

This yields a plasmon velocity of the form:
\begin{equation*}
    v_{\mathrm{1D}} \; = \;  \sqrt{\frac{eWG}{m^*C\mu}} \;,
\end{equation*}
where $e$ is the elementary charge, $m^*$ is the effective electron mass, $C$ is the capacitance per unit of length set by the electrostatic gates and $\mu$ is the mobility. 
In the explored conductance range (2–10 $G_0$), the geometrical parameters remain essentially unchanged. 
The velocity dependence is therefore dominated by the conductance, leading to $v_{\mathrm{1D}} \propto \sqrt{G}$.

\begin{figure}[!h]
    \centering
    \includegraphics[scale = 1.0]{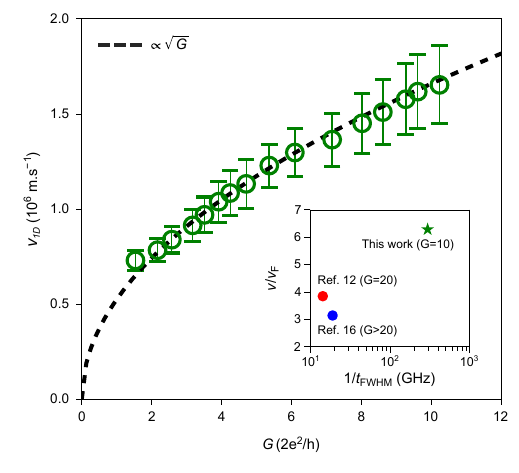}
    \caption{
    \textbf{Plasmonic velocity in 1D electronic waveguide.} 
    The main figure is the \SI{}{\tera\hertz} plasmonic velocity as a function of the conductance $G$ in 1D electronic waveguide. 
    The dotted line is a fit to the experimental data following a $\sqrt{G}$ dependence \cite{Kukushkin2005}. 
    The inset presents the frequency-domain figure of merit of the plasmonic velocity, comparing the present work with previously reported experimental studies.  
    }
    \label{fig4}
\end{figure}

To put our results in context, the inset of Fig. \ref{fig4} compares the plasmon velocities of our measurements with previously reported values for one-dimensional electronic waveguides \cite{Takada2025,Roussely2018}. 
In those studies, voltage pulses of \SI{52}{\pico\second} (\SI{19}{\giga\hertz}) and \SI{70}{\pico\second} (\SI{14}{\giga\hertz}) were used. 
By contrast, our experiment employs ultra-short pulses of just \SI{4.3}{\pico\second} ($\approx\,$\SI{230}{\giga\hertz}), more than 15 times shorter.
This strong reduction in pulse width leads to a significantly higher plasmon velocity in our device compared to earlier measurements.
This can be rationalised through two complementary arguments.
First, higher frequencies correspond to a reduced spatial extent of the excitation, which modifies the local electric field distribution and contributes to an increased propagation velocity.
Second, at higher frequencies the plasmonic excitation is less efficiently screened by the electrostatic gates, resulting in velocities approaching \SI{1e7}{\meter\per\second}, as reported for ungated GHz one-dimensional plasmonic modes \cite{Kukushkin2005, Kumada2011}.

\section*{Discussion and Conclusion}

We have demonstrated an integrated THz quantum nanoelectronic platform capable of generating, guiding and detecting plasmonic excitations in a one-dimensional electronic waveguide. 
By combining ultrafast optoelectronic photoconductive switches with a high-mobility GaAs/AlGaAs quantum device, we achieved the direct injection of electrical wavepackets with temporal durations down to \SI{4}{\pico\second} into an engineered quantum conductor. 
Time-resolved measurements reveal clear signatures of one-dimensional plasmon propagation, while we further establish active electrostatic control of the plasmon velocity through a tunable confinement of the plasmonic excitations in the electronic waveguide.

Compared with previous experiments operating in the \SI{}{\giga\hertz} regime, our work pushes plasmonic transport into the \SI{}{\tera\hertz} domain, increasing the operational bandwidth by more than an order of magnitude. 
In this regime, the spatial extent of the electronic wavepacket becomes much smaller than the dimensions of the quantum device itself, opening access to a new regime of  quantum transport. 
Recent theoretical proposals have predicted interference protocols capable of probing the internal dynamics of quantum systems on ultrafast timescales \cite{Gaury2014,Kloss2025a,Saha2025}, but their implementation requires electronic excitations compressed to only a few picoseconds. Our platform provides precisely this capability.

Beyond establishing a new frontier for ultrafast electron quantum optics, these results represent a major step towards scalable quantum architectures based on flying electronic excitations \cite{Edlbauer2022, Pomaranski2024, Pomaranski2025}. 
In particular, emerging quantum computing schemes rely on coherent quantum loops in which ultrashort electronic excitations circulate while being dynamically manipulated through successive quantum operations \cite{Takeda2017}. 
The ability demonstrated here to generate, guide and control picosecond plasmonic wavepackets in engineered ballistic conductors brings such architectures within experimental reach. More broadly, THz quantum nanoelectronics provides a powerful platform for investigating non-equilibrium many-body dynamics, ultrafast quantum interference phenomena, and future solid-state quantum information processors operating at unprecedented speeds.

\section*{Methods}
\subsection*{Device fabrication}

The device was fabricated from a GaAs/AlGaAs heterostructure hosting a two-dimensional electron gas (2DEG) located \SI{100}{\nano\meter} below the surface. 
The 2DEG electron density was \SI{1.86e11}{\per\centi\meter\squared} in the dark and \SI{3.66e11}{\per\centi\meter\squared} under illumination, with corresponding mobilities of \SI{5.2e5}{\centi\meter\squared\per\volt\per\second} and \SI{1.24e6}{\centi\meter\squared\per\volt\per\second}, respectively. 
To monolithically integrate the THz opto-electronics with the quantum device, a \SI{1}{\micro\meter}-thick low-temperature-grown GaAs (LT-GaAs) was deposited at \SI{300}{\degreeCelsius} via molecular beam epitaxy far below the surface,
and separated from the 2DEG by a \SI{500}{\nano\meter}-thick GaAs and a \SI{10}{\nano\meter}-thick AlAs sacrificial layer (Fig. \ref{fig1}a and Supplementary Fig. S1a). 
The thick GaAs buffer ensured a high-quality crystalline region supporting the formation of a high-mobility 2DEG, while the embedded LT-GaAs layer provided a compact integration of electronic and photoconductive functionalities.

The mesa structure was defined by a two-step etching process. GaAs was selectively etched using a citric acid solution, followed by selective removal of the AlAs sacrificial layer using HCl. This process created a staircase geometry enabling the coplanar waveguide (CPW) to bridge the \SI{610}{\nano\meter} height difference towards the ohmic contacts. The electronic waveguide consisted of Ti/Au electrostatic gates defined by electron-beam lithography. The photoconductive switches, CPW, and interconnects to the electrostatic gates were fabricated from Ti/Au using photolithography. Ohmic contacts were formed by deposition of Ni/Au/Ge/Ni, followed by annealing at \SI{370}{\degreeCelsius} for \SI{1}{\minute} and \SI{430}{\degreeCelsius} for \SI{2}{\minute} under a continuous forming gas flow of \SI{850}{\centi\metre^{3}\per\minute} (\SI{5}{\percent} H\textsubscript{2} in Ar).
The electronic waveguide of the quantum device has a lithographically defined width W of \SI{500}{\nano\meter}. 
When a one-dimensional channel is formed by electrostatic gate depletion, the effective width is reduced to $\approx$ \SI{300}{\nano\meter} as the depletion length is of the same order as the depth of the 2DEG ($\approx$\SI{100}{\nano\meter}).

\subsection*{On-chip THz measurement in dilution refrigerator}

The experimental setup consisted of dispersion-compensated optical fibres installed in a dilution refrigerator. 
At the cold-finger stage, a set of lenses focused the optical pulses onto the sample, where the photoconductive switches (PCSs) were located. 
Since the optical spots were fixed in space, the sample was mounted on piezoelectric nanopositioners (Attocube Systems GmbH) providing motion along the x, y, z, and rotational axes to align the PCSs with the optical beams.

Optical excitation was provided by a \SI{100}{\femto\second} pulsed laser (Menlo Systems GmbH) operating at a wavelength of \SI{780}{\nano\meter} with a repetition rate of \SI{250}{\mega\hertz}. 
The laser beam was split into two paths, one of which passed through a mechanical delay line and an optical chopper operating at \SI{329}{\hertz} for lock-in detection of the THz signal. 
Before coupling into the fibres and the dilution refrigerator, both beams were pre-compensated for fibre dispersion using a GRISM compressor.

\section*{Data availability}

The experimental data generated in this study have been deposited in the Zenodo database under accession code XXXXXXXX

\section*{Acknowledgments}

The sample was fabricated in the clean room facility of Neel Institute (Grenoble) and we thank all the clean room staff for their assistance.
This project has received funding from the European Union H2020 research and innovation program under grant agreement No. 862683, “UltraFastNano" as well as from the European Innovation Council and SMEs Executive Agency under Grant Agreement No. 101185712 "ELEQUANT".
C.B., H.S., J-F. R, P.B.V., N.R., K.B. and J.S. acknowledge funding from the Agence Nationale de la Recherche under the France 2030 programme, reference ANR-22-PETQ-0012. 
C.B acknowledges financial support form the European Research Coucil under Grant Agreement ID 101201077 "UltraWave".
C.B., J-F. R., P.B.V., G.G. and K.B. acknowledge funding from the Agence Nationale de la Recherche, references "QTERA" ANR-15-CE24-0007 and "STEPforQubits" ANR-19-CE47-0005.
M.A. acknowledges the MSCA co-fund QuanG Grant No. 101081458, funded by the European Union and the program QuanTEdu-France n° ANR-22-CMAS-0001 France 2030.
L.M. acknowledges the program QuanTEdu-France n° ANR-22-CMAS-0001 France 2030.
T.V. acknowledges funding from the French Laboratory of Excellence project "LANEF" (ANR-10-LABX-0051).
G.G. acknowledges EPSRC ``QUANTERAN" (grant number EP/X013456/1) and Royal Society of Edinburgh ``TEQNO" and ``TFLYQ'' (grant numbers 3946 and 4504).
A.D.W. and A.L. thank the DFG via ML4Q EXC 2004/1 - 390534769, the BMBF-QR.X Project 16KISQ009 and the DFH/UFA Project CDFA-05-06.

Views and opinions expressed are those of the author(s) only and do not necessarily reflect those of the European Union or the granting authority. Neither the European Union nor the granting authority can be held responsible for them.\\

\section*{Author contributions}
T.V. designed and fabricated the sample. T.V. and U.A. performed the measurements with support from C.G. L.M. M.A. S.O. J.S..
C.G. and G.G. developed the experimental setup.
T.V. and U.A. analysed the data.
A.L. and A.D.W provided the high-mobility GaAs/AlGaAs heterostructure.
N.R and K.B performed the time resolved measurements at 300 K under the supervision of J-F.R. and P-B.V..
T.V., U.A., G.G. and C.B wrote the manuscript with the feedback of all the authors.
G.G., J-F.R, P-B.V. and C.B. supervised the experimental work.
C.B. has initiated the project.

\section*{Competing financial interests}

The authors declare no competing financial interests.

\clearpage
\newpage
\onecolumngrid
\setcounter{figure}{0}
\setcounter{equation}{0}
\renewcommand{\thefigure}{S\arabic{figure}}
\begin{center}

{\Large\bfseries
Supplementary material: Active control of THz plasmon propagation in a
one-dimensional electronic waveguide
}

\vspace{1em}

Thomas Vasselon$^{1}$,
Uzer Ahmad$^{1}$,
Clément Geffroy$^{1}$,
Lucas Mazzella$^{1}$,
Matteo Aluffi$^{1}$,
Seddik Ouacel$^{1}$,
Jashwanth Shaju$^{1}$,
Kevin Bredillet$^{2}$,
Nathan Roussel$^{2}$,
Jean-François Roux$^{2}$,
Pierre-Baptiste Vigneron$^{2}$,
Arne Ludwig$^{3}$,
Andreas D. Wieck$^{3}$,
Matias Urdampilleta$^{1}$,
Hermann Sellier$^{1}$,
Giorgos Georgiou$^{4}$,
Christopher B\"auerle$^{1}$

\vspace{0.5em}

{\small
$^{1}$Université Grenoble Alpes, CNRS, Grenoble INP, Institut Néel,
Grenoble, 38000, France

$^{2}$Université Grenoble Alpes, Université Savoie Mont-Blanc,
CNRS, Grenoble INP, CROMA, Grenoble, 38000, France

$^{3}$Lehrstuhl für Angewandte Festkörperphysik,
Ruhr-Universität Bochum, Bochum, Germany

$^{4}$James Watt School of Engineering, Electronics and Nanoscale Engineering,
University of Glasgow, Glasgow, G12 8QQ, United Kingdom
}

\vspace{0.5em}

{\small Corresponding author: christopher.bauerle@neel.cnrs.fr}

\end{center}


\section{Quantum -- THz electronic circuit and experimental setup}

\subsection{Nanofabrication of the device}

\noindent The quantum device was fabricated from a dedicated heterostructure which comprises a high mobility 2DEG located at a depth of approximately 100 nm as well as a monolithically integrated opto-electronic material, a layer of low-temperature-grown GaAs (LT-GaAs) positioned at a depth of 610 nm and with thickness of \SI{1}{\micro\metre}.

\begin{figure}[h]
    \centering
    \includegraphics[scale = 0.8
]{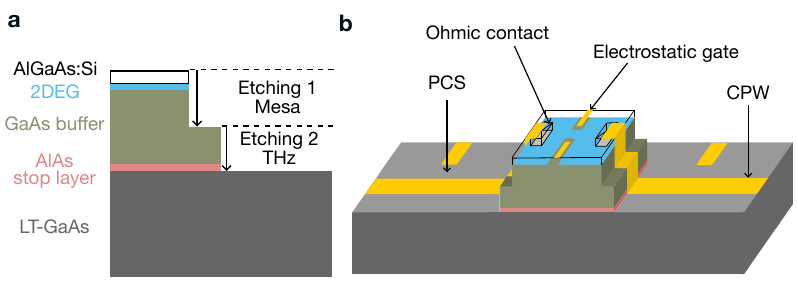}
    \caption{
    \textbf{Schematic of the heterostructure.} 
    \textbf{a.} Cross-sectional view of the wafer highlighting the two etching steps. 
    The first step defines the mesa, while the second exposes the LT-GaAs surface for THz circuit fabrication. 
    \textbf{b.} Schematic of a complete THz device, including the patterned THz circuitry comprising the CPW and the PCS. 
    On the mesa, ohmic contacts are deposited to provide electrical access to the 2DEG, and electrostatic gates are patterned to define the desired quantum circuits.
    }
    \label{nanofab}
\end{figure}

This material integration is needed to combine THz generation using photoconductive switches with a high mobility 2DEG on the same chip. 
The resulting two dimensional electron gas has a carrier density of \SI{1.86e11}{\per\centi\meter\squared} (\SI{3.66e11}{\per\centi\meter\squared}) and a mobility \SI{5.2e5}{\centi\meter\squared\per\volt\per\second} (\SI{1.24e6}{\centi\meter\squared\per\volt\per\second}) in dark (under illumination).
The LT-GaAs layer shows a 500 fs electron trapping time
as measured by time-resolved reflectometry.
The nanofabrication process consists of several steps which we summarize in the following: First, a two-step selective wet etching process forms the mesa structure and exposes the active layers, creating a staircase profile that allows electrical connection between different device regions (see Fig. \ref{nanofab}).

Ohmic contacts are then fabricated using an Au/Ni/Ge stack followed by thermal annealing, providing electrical contact to the 2DEG.
Electron-beam lithography is used to define the nanoscale electrostatic gates forming the quantum circuits, followed by Ti/Au deposition. 
Larger-scale  electrical connections, including photoconductive switches and coplanar waveguides, are then patterned using UV lithography and metal deposition.

\subsection{Quantum device}

\begin{figure}[h]
    \centering
    \includegraphics[scale = 1.0]{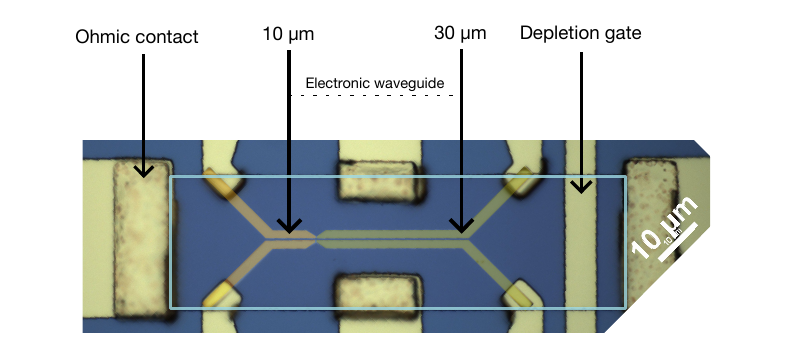}
    \caption{
    \textbf{Optical image of the mesa.}
    The blue square indicates the etched mesa containing the 2DEG. 
    All relevant components are labelled: the four ohmic contacts, electronic waveguides of 10 and \SI{30}{\micro\metre}, and the depletion gate.
    }
    \label{device}
\end{figure}

\noindent Figure \ref{device} shows an optical microscope image of the full quantum device. 
In addition to the ohmic contacts and electrostatic gates, a global depletion gate was implemented, consisting of a 5-µm-wide Ti/Au electrode fabricated simultaneously with the THz circuits.

This gate serves two purposes: (i) it enables the reproduction of gated plasmon cavity experiments reported previously \cite{Wu2015}, and (ii) it allows for the discrimination betweenthe plasmonic signals within the 2DEG and substrate-mediated contributions, by depleting the 2DEG and suppressing electrical conduction.

\subsection{THz electronic circuit}

\noindent To guide the THz pulse towards the ohmic contact of the 2DEG we used coplanar waveguides (CPW) patterned onto the GaAs heterostructure.  The CPW was designed for a characteristic impedance of 50 $\Omega$, using a \SI{30}{\micro\metre} signal track separated by \SI{20}{\micro\metre} gaps from the ground planes.

\begin{figure}[h]
    \centering
    \includegraphics[scale = 1.0]{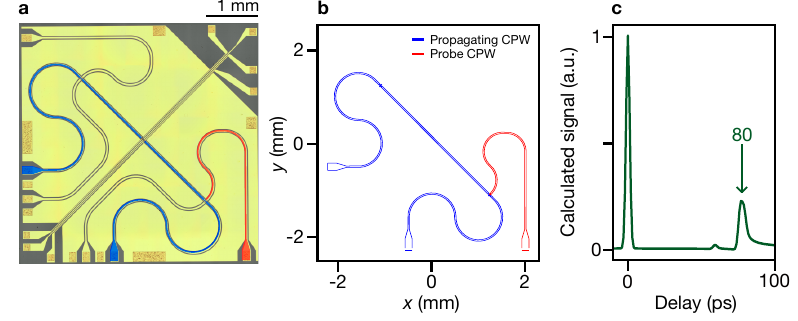}
    \caption{
    \textbf{Design of the THz circuit.}
    \textbf{a.} Optical image of the sample containing the coplanar waveguides (CPW) and photoconductive switches. 
    The CPW along which the THz signal propagates is highlighted in blue, and the probe CPW is shown in red.
    \textbf{b.} Extracted CPWs used for simulations. 
    For this simulation the 2DEG has been removed and only a continuous CPW has been taken into account.
    \textbf{c.} Simulated THz signal corresponding to the CPWs shown in \textbf{b}. 
    The first reflection, arising from the bonding pad, occurs at \SI{80}{\pico\second}. 
    }  
    \label{simu}
\end{figure}

To study THz pulse propagation in time-resolved measurements, parasitic reflections must be minimized or delayed beyond the time window relevant for signal propagation through the quantum device. The dominant source of reflections is the impedance mismatch introduced by the wire bonds connecting the coplanar waveguide (CPW) to the external circuitry. 
To mitigate this effect, the THz circuit was designed to maximize the distance between the photoconductive switches (PCS) and the tapered bonding pads, thereby increasing the delay before the first reflected pulse returns to the measurement PCS. 
This design was constrained by the chip dimensions (< 5\,\rm{mm} × 5\,$\rm{mm}$) and the fixed separation between the two laser spots, requiring a long CPW with several smooth bends, with  bend radii are larger than \SI{500}{\micro\metre}, as shown in Fig.\,\ref{simu}a.

\begin{figure}[h]
    \centering
    \includegraphics[scale = 1.0]{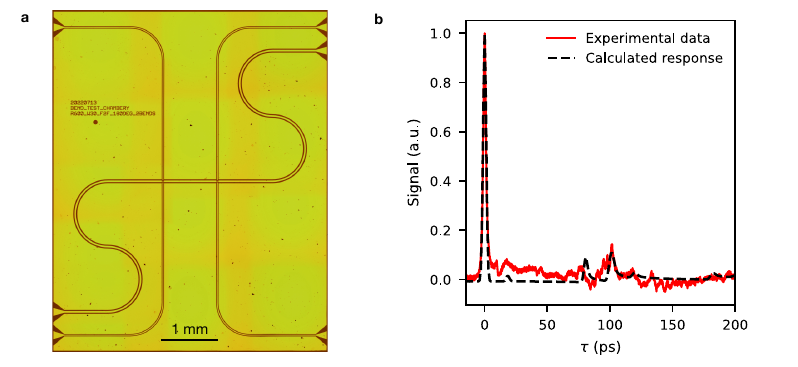}
    \caption{
    \textbf{Reflection-free bending CPW.}
    \textbf{a.} Optical image of a THz circuit fabricated from an LT-GaAs wafer.
    It consists of a bent CPW with a radius of curvature of \SI{600}{\micro\meter} and a distance of \SI{1}{\milli\meter} between the two PCS.
    \textbf{b.} Experimental data compared with the response calculated using our home-built numerical toolkit.
    The solid red curve shows a pump-probe measurement performed at \SI{300}{\kelvin}, while the dashed black curve corresponds to the calculated response.
    Excellent agreement is observed, with reflections appearing at approximately \SI{100}{\pico\second}.
    }
    \label{data_R600}
\end{figure}

Since sharp bends introduce additional impedance mismatches and reflections, the CPW geometry was systematically optimized by varying the bend radius and exploring alternative layouts, including spiral-shaped waveguides that provide more gradual curvature transitions. Experimental characterization established that bend radii larger than \SI{500}{\micro\metre} effectively suppress bend-induced reflections. Representative measurements for a bend radius of \SI{600}{\micro\metre} are shown in Fig.\,\ref{data_R600}.

To quantify the remaining reflections, we developed a home-made numerical tool that calculates time-domain signal propagation through the complete THz circuit based on the device GDS layout (Fig.\,\ref{simu}a,\,b), while excluding the quantum device itself (Fig.\,\ref{simu}c and Fig.\,\ref{data_R600}b). Combined with the experimentally identified reflection points, the simulations show that, for bend radii larger than \SI{500}{\micro\metre}, the residual reflections originate predominantly from the bonding pads. The first reflected pulse is predicted to return after approximately 80 ps as shown in Fig.\,\ref{simu}c, consistent with the propagation delay associated with the distance between the PCS and the bonding pads. This defines the effective temporal window available for time-resolved measurements before the onset of parasitic reflections.

\subsection{Experimental setup}

\noindent A femtosecond laser (C-Fiber 780, Menlo Systems) delivering 100 fs pulses at a repetition rate of 250 MHz and a wavelength of 780 nm is used to generate electron–hole pairs in LT-GaAs. 
The pulse train is first sent through a GRISM compressor to pre-compensate for dispersion in the optical fibers. 
It is then split into pump and probe arms, the latter passing through a mechanical delay line and an optical chopper, to enable time-resolved measurements and lock-in detection.

\begin{figure}[h]
    \centering
    \includegraphics[scale = 1.0]{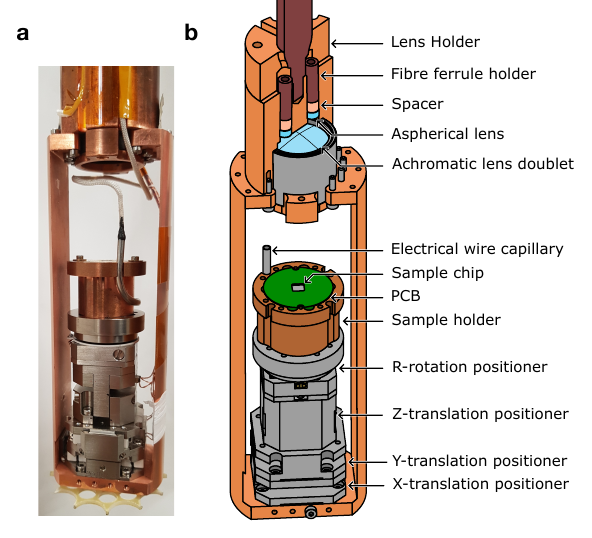}
    \caption{
    \textbf{Coldfinger of the experimental setup.}
    \textbf{a.} Photograph of the coldfinger comprising the lens as well as the sample stage mounted on XYZR piezoelectric motors used to align the photoconductive switches with respect to the fs laser beams. 
    \textbf{b.} 3D technical rendering of the cold finger, with the lens block partially exposed.
    }  
    \label{coldfinger}
\end{figure}

Both arms include variable attenuators (half-wave plate and polarizer) to control the optical power before coupling into fibers at room temperature. 
The pulses are guided through optical fibers into the dilution refrigerator, where four fibers connect from a room temperature connector to the mixing chamber stage (only two are used here). 
A custom made optical lens system focuses the pulses onto the photoconductive switches placed on the sample, as shown in Fig. \ref{coldfinger}. 
The separation between the two optical beams (pump/probe) is fixed by the focussing lens system, and is 3.3 mm.

\newpage

\section{Device characterization}

\subsection{Transport characterization of the quantum device}

\begin{figure}[!h]
    \centering
    \includegraphics[scale = 1.00]{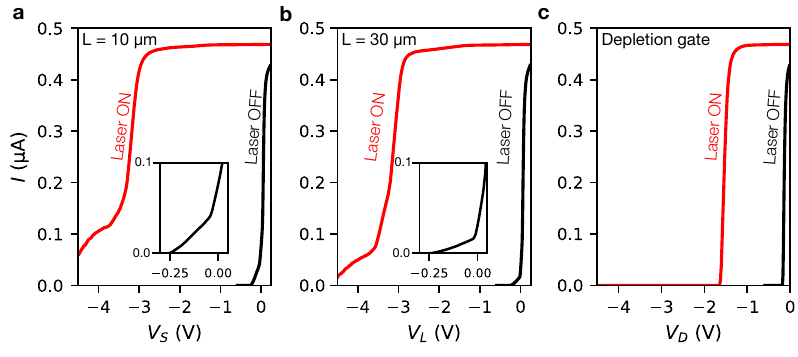}
    \caption{
    \textbf{Quantum transport characterisation of the quantum device before and after illumination.}
    For all panels, black curves correspond to measurements without laser illumination, and red curves correspond to measurements under laser illumination.
    Conductance measurements  when depleting the electron gas applying negative voltages to the two electrostatic gates of the \textbf{a.} \SI{10}{\micro\metre} electronic waveguide.
    \textbf{b.} \SI{30}{\micro\metre} electronic waveguide.
    \textbf{c.} large depletion gate.
    The insets in panels \textbf{a} and \textbf{b}, demonstrate a closer view of the conductance under dark conditions. 
    In this regime, the electronic waveguides are fully formed and the system behaves as a quasi-1D channel.
    }  
    \label{PO}
\end{figure}

\noindent Transport measurements are performed by biasing the electrostatic gates with negative gate voltages,
thereby depleting the electron density in the underlying 2DEG. Fig. \ref{PO} shows the
two-terminal conductance of the short (S) and long (L) electronic waveguides, as well
as the depletion gate (D), before (black curves) and after illumination (red curves).
The device has been bias cooled  by applying +0.2 V on all electrostatic gates, except the depletion gate.

We observe a typical pinch-off curve, characterized by a rapid decrease in conductance around \SI{-0.2}{\volt} (laser off) due to the depletion of the 2DEG underneath large gate structures. 
The slope change observed for lower voltages is a signature of the formation of a quasi-one-dimensional channel.

After illumination, the pinch-off curves shift towards more negative gate voltages, indicating an increase in the carrier density of the 2DEG. 
Illumination at \SI{780}{\nano\meter} excites the DX centres, leading to an increase in both the carrier density and the mobility of the 2DEG \cite{Mooney1987,Hayne1996}. At \SI{4}{\kelvin}, this photo-induced state persists over experimental timescales, and the sample must be heated above \SI{100}{\kelvin} to reset the DX centres.
To avoid exceeding the breakdown voltage of the Schottky barrier, we limit the gate voltage sweep to \SI{-4.5}{\volt}.

\subsection{Electrical characterization of photo-conductive switches}

\begin{figure}[!h]
    \centering
    \includegraphics[scale = 0.8]{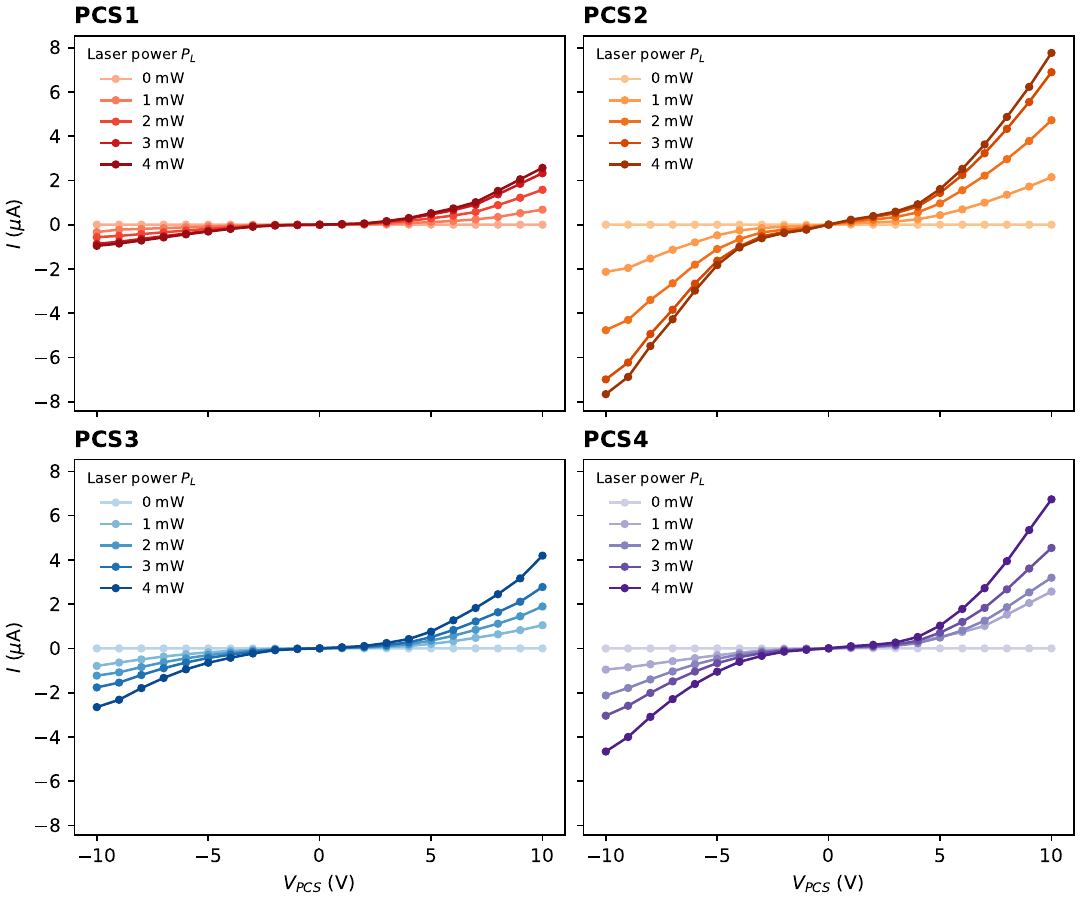}
    \caption{
    \textbf{IV characteristics of the four photoconductive switches.}
    For each photoconductive switch, the I-V characteristics were measured under different laser powers (P$_\mathrm{L}$) varying from  0 to \SI{4}{\milli\watt}.
    }
    \label{PCS}
\end{figure}

\noindent To characterize the photoconductive switches (PCSs), we measured their current–voltage (IV) characteristics at a temperature of \SI{4}{K}.
This first required aligning the optical laser spots onto the PCSs. The IV measurements (shown in Fig.~\ref{PCS}) were performed by biasing the PCS and collecting the resulting current through the coplanar waveguide's center conductor while varying the optical laser power from 0 to \SI{4}{\milli\watt}.
The photocurrent increases with both the optical power and the applied bias voltage.
The PCSs exhibit different efficiencies, reflected by variations in the photocurrent for a given optical power (P$_\mathrm{L}$) and bias voltage (V$_\mathrm{PCS}$). All switches display the expected photoconductive behaviour, with the photocurrent dropping to nearly \SI{0}{\micro\ampere} in the dark (\SI{0}{\milli\watt} illumination).

\newpage

\begin{figure}[!h]
    \centering
    \includegraphics[scale = 0.85]{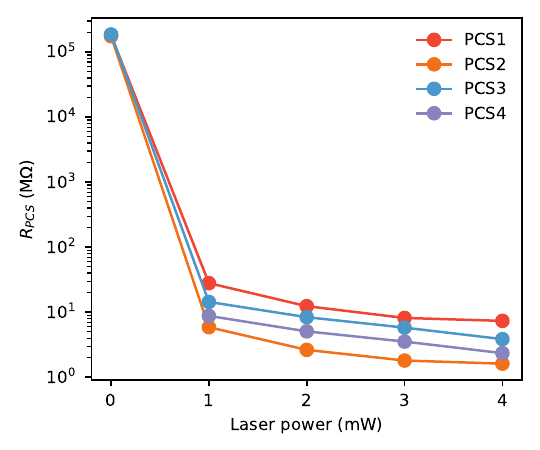}
    \caption{
    \textbf{Mean resistance value of the photoconductive switches as a function of the laser power.}
    Each photoconductive switch exhibits a similar trend under illumination with a strong decrease in  resistance upon illumination.
    }
    \label{R_vs_pow}
\end{figure}

\newpage
From these measurements, the mean resistance of each PCS was extracted as a function of the optical power, as shown in Fig.~\ref{R_vs_pow}, by averaging over all point of the non-linear IV curve.
Without any laser illumination (dark), the resistance reaches values on the order of several hundred G$\Omega$, while under illumination it decreases to the M$\Omega$ range. This change of 4 to 5 orders of magnitude confirms the efficient switching behaviour required for pump–probe measurements.

\newpage

\subsection{Pump-probe experiments}

\begin{figure}[!h]
    \centering
    \includegraphics[scale = 1.0]{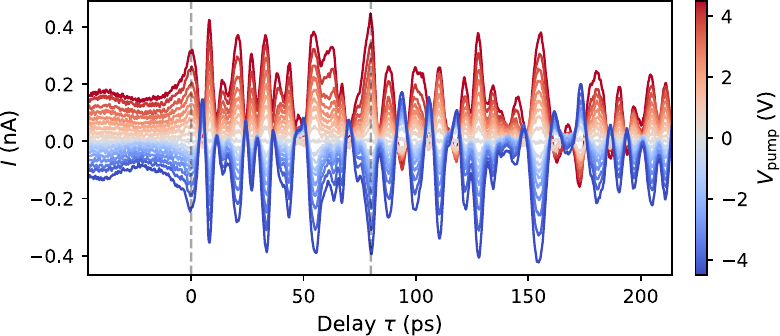}
    \caption{
    \textbf{Time-resolved measurements of the transmitted THz signal for different bias voltages applied to the pump photoconductive switches}.
    V$_{\mathrm{PCS1}}$ and V$_{\mathrm{PCS2}}$ are swept from \SI{-4.5}{\volt} to \SI{4.5}{\volt}, while the optical power is fixed at \SI{3.7}{\milli\watt} on the probe and \SI{2.6}{\milli\watt} on the pump.
    }
    \label{TR_pump}
\end{figure}

\noindent Time-resolved measurements are performed using the pump-probe technique.
The probe beam is mechanically modulated at 329 Hz
for lock-in detection. Here we focus on the transmitted signal through
the device, without applying any gate voltage to the electrostatic gates of the quantum circuit. Fig. \ref{TR_pump} shows the transmitted
signal for different V$_{pump}$ (simultaneously applied to PCS1 and PCS2), allowing
us to verify the linear dependence of the signal on the pump bias. 
Our measurements show that the signal does not consist of a single transmission peak, but
rather a complex but very reproducible plasmonic response as the picosecond electrical pulse propagated through the device. 
At $\tau$ = 80 ps (grey dashed line in Fig. \ref{TR_pump}), the signal amplitude shows a small increase, which we attribute to the reflected pulse after a full round trip, and which propagates through the quantum device for a second time, in agreement with the simulations shown in Fig. \ref{simu}. 
As pointed out in the main text, this behaviour which we call
a background signal contribution to our quantum transport measurements, likely arises from direct THz transmission through the semiconductor substrate beneath the 2DEG, or due to a possible capacitive coupling between the CPW transmission lines on either side of the mesa.

\subsection{Sliding contact technique}

\begin{figure}[h]
    \centering
    \includegraphics[scale = 1]{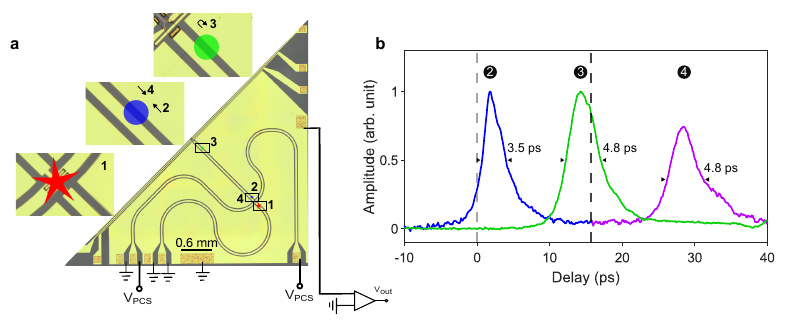}
    \caption{
    \textbf{a. Optical microscope image  of the device}. It  comprising the coplanar waveguide (CPW) and the integrated PCS. The two probing positions used for the sliding contact measurements are indicated by the blue and green markers.
    \textbf{b.} Corresponding time-domain THz waveforms measured at positions 2 and 3, respectively close to the generation PCS and to the ohmic contact. 2 is the forward propagating pulse coming from position 1; 4 is the backward propagating pulse coming from the ohmic contact. 
    We measure a mean group velocity $\langle v_g \rangle \approx 102$ µm.ps$^{-1}$. This yields an arrival time at the ohmic contact of $t = 15.7~\text{ps}$, indicated by a dashed vertical line on the graph. 
    The pulse width measured at position 3 has a full width at half maximum (FWHM) of 4.8 ps. This value is however overestimated, as the forward-propagating pulse partially overlaps with the reflected pulse returning from the ohmic contact.
    }
    \label{sliding_contact}
\end{figure}

\noindent To characterize the excitation pulse prior to its arrival at the ohmic contact, we employ the sliding-contact technique \cite{Eusebe2005} using a free-space optical THz setup operating at 300 K and an identical sample as the one studied in the main manuscript. 
This approach enables the reconstruction of the pulse waveform at arbitrary positions along its propagation path. 
In contrast to the pump–probe configuration described above, which relies on two photoconductive switches (PCS), the probing is performed by optically exciting the gap between the central conductor and one of the ground planes.
In this configuration, the ground plane is connected to the lock-in amplifier, as illustrated in Fig. \ref{sliding_contact}. Similarly to the pump-probe experiment carried out at 4K, we promote the excitation of the even coplanar waveguide (CPW) mode by symmetrically biasing PCS1 and PCS2.

Two distinct positions along the central conductor, designated 2 and 3 in Fig. \ref{sliding_contact}, are selected to characterize the THz pulse close to the generation PCS (position 2) and close to the ohmic contact (position 3) , respectively. The location of the generation PCS is indicated by the position 1. 
At each position, the inter-conductor gap is illuminated by  the probe laser pulse with an optical power of 10 mW. 
A transient photocurrent is generated when the THz pulse traverses the illuminated region, thereby enabling time-resolved sampling of the waveform.
In this experiment, both laser pulses are mechanically modulated at two different frequencies f$_1$ and f$_2$, the signal is then detected at the sum frequency using a lock-in amplifier. 

The corresponding time-domain THz waveforms measured at
positions 2 and 3 are shown in Fig. \ref{sliding_contact}b. The traces have been plotted on a common time equivalent scale where the zero position is the estimated time at which the pulse is generated at position 1. One can observe three different pulses named 2, 3 and 4 in Fig. \ref{sliding_contact}b.
Pulse 2 (blue curve in Fig. \ref{sliding_contact}b) is measured at position 2 after a propagation time of 1.8 ps after pulse generation at position 1 while propagating towards the ohmic contact. At the input of the ohmic contact, the pulse is partially reflected back because of impedance mismatch in between the ohmic contact and the transmission line, while the other part of the pulse is  injected into the ohmic contact and excites the 2DEG structure (see main text). As a consequence,  pulse 3 (green curve in Fig. \ref{sliding_contact}b), which is measured very close to the ohmic contact, corresponds to a superposition of the wave packet propagating towards the ohmic contact and part of this wave packet that is reflected back at the ohmic contact input. Finally, pulse 4 (purple curve in Fig. \ref{sliding_contact}b) is the pulse detected at position 2 after its partial reflection from the ohmic contact and propagation back towards position 1.
Overall, we observe that, as the pulse propagates, it broadens while its amplitude decreases due to propagation losses and group-velocity dispersion in the transmission line.

In this coplanar waveguide (CPW) the quasi-transverse electromagnetic  mode (quasi-TEM) is the dominant propagation mode for frequencies below the cutoff frequency $f_{\text{TE}} \approx 43.4 ~ \text{GHz}$ \cite{alma991008107804806161}. 
The propagation of these lower spectral components of the pulse is mostly affected by conductor losses and dielectric losses \cite{alma991008108405806161}. 
However, for higher frequencies, the quasi-TEM mode progressively couples to the TE substrate mode through leaky-mode radiation \cite{Potts2023, FrankelM.Y.1991Taad}. This coupling process leads to radiation losses that increase as a power law of the frequency. Consequently, the pulse experiences the temporal distortion and attenuation observed in Fig. \ref{sliding_contact}b.

By analysis of the temporal positions of the recorded pulses 2 and 4, we determine a mean group velocity of $102 ~ \text{µm} \cdot \text{ps}^{-1}$ in good agreement with the theoretical values extrapolated from \cite{HasnainG.1986DoPP}. Finally, in order to determine the FWHM of the pulse that enters the ohmic contact, we fitted the set of measurements of the FWHM of pulses 2 and 4 using a power law and extrapolated a pulse width near the ohmic contact ($1600~\text{µm}$ from the PCS) of $\tau \approx 4.3~\text{ps}$ in good agreement with the $\tau \approx 4.8~\text{ps}$ FWHM of pulse 3 detected close to the ohmic contact.

\newpage

\section{Plasmonic excitation - extended information.}

\subsection{Two-dimensional plasmonic excitation}

\begin{figure}[h]
    \centering
    \includegraphics[scale = 1.0]{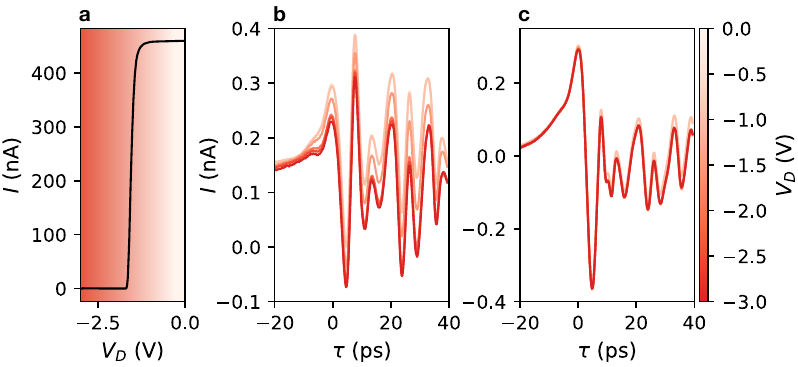}
    \caption{
    \textbf{Plasmonic excitation from even and odd modes.}
    \textbf{a.} Transmitted current as a function of the depletion gate.
    The red colour gradient matches the colour scale used for the time-resolved measurements on the right.
    \textbf{b.} Time-resolved measurements for predominantly even-mode excitation, obtained with $V_{\mathrm{PCS1}}$ = $V_{\mathrm{PCS2}}$.
    \textbf{c.} Time-resolved measurements for predominantly odd-mode excitation, obtained with $V_{\mathrm{PCS1}}$ = - $V_{\mathrm{PCS2}}$.
    }
    \label{odd_vs_even}
\end{figure}

\noindent To investigate the two-dimensional plasmonic excitations, time-resolved measurements of the transmitted signal were first performed using only the depletion gate, shown in Fig.\ref{device}. 
Figure \ref{odd_vs_even}a presents the pinch-off characteristic of this gate.
To study how the excitation of 2D plasmons is affected by the CPW mode, we inject THz pulses into the even (coplanar) or odd mode (slotline) of the CPW, by biasing both sides of the photoconductive switches. 
Even modes are generated using symmetric biasing of the photoconductive switches, corresponding to $V_{\mathrm{PCS1}} = V_{\mathrm{PCS2}}$. 
In contrast, odd modes are predominantly excited using antisymmetric biasing, $V_{\mathrm{PCS1}} = -V_{\mathrm{PCS2}}$.

For even-mode excitation, Fig.\ref{odd_vs_even}b, the transmitted current decreases progressively as the depletion gate is pinched off. 
In contrast, the signal remains nearly unchanged for odd-mode excitation (Fig.\ref{odd_vs_even}c). 
This behaviour demonstrates that the even THz mode is more efficiently coupled to the 2DEG, consistent with previous reports \cite{Wu2015}.

For both excitation configurations, a residual transmitted signal remains visible for depletion voltages below \SI{-1.7}{\volt}, despite the 2DEG being electrically disconnected beyond this gate voltage. 
This behaviour is similar to that reported in Ref. \cite{Wu2015}, where a background contribution originating from direct THz transmission through the mesa was observed.
In our device, this parasitic signal may originate from the vertical transition of the coplanar waveguide (CPW) before reaching the mesa, effectively creating a capacitive coupling across the quantum circuit.

\begin{figure}[!h]
    \includegraphics[scale = 0.8]{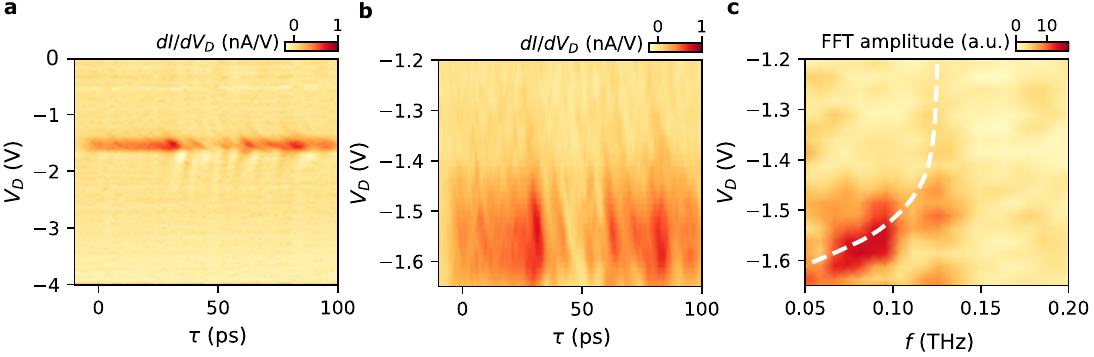} 
    \caption{
    \textbf{Plasmonic resonance of a gated cavity defined by the depletion gate.}
    \textbf{a.} Two-dimensional map of the derivative of the time-resolved signal with respect to the voltage $V_D$ applied to the depletion gate.
    \textbf{b.} Zoom of panel \textbf{a}, highlighting the abrupt change in current as the carrier density decreases in the gated region (see \textbf{a}).
    \textbf{c.} Fast Fourier Transform (FFT) of panel \textbf{b}. The white dashed line corresponds to the analytical calculation for the first order gated plasmon resonance in the cavity defined by the depletion gate.
    }
    \label{gated_plasmon}
\end{figure}

The depletion gate is then used to define a gated plasmonic cavity, similarly to the study reported in Ref. \cite{Wu2015}. This demonstrates the presence of a plasmonic resonance in our device.
To isolate the plasmonic contribution from the background signal, the time-resolved traces are numerically differentiated with respect to the gate voltage $V_D$, suppressing contributions originating outside the 2DEG. 
Fig. \ref{gated_plasmon}a shows the resulting map of $dI/dV_D$ as a function of delay time and depletion voltage.
A non-zero signal is observed when the carrier density inside the cavity changes, corresponding to the decrease in current observed in the two-terminal transport measurement shown in Fig.\ref{odd_vs_even}a. 
A closer look of this transition region is presented in Fig.\ref{gated_plasmon}b.
The Fast Fourier Transform (FFT) of these measurements, shown in Fig. \ref{gated_plasmon}c, reveals a confined plasmonic mode inside the cavity, with a characteristic frequency that decreases as the gate voltage becomes more negative.

The cavity supports discrete plasmonic modes determined by its length. 
The allowed plasmon wavenumbers satisfy the resonance condition $k = \frac{N\pi}{L}$, where N is the mode index and L is the cavity length.
The plasmon dispersion relation in a two-dimensional electron gas is given by:
\begin{equation}
    \omega_P = \sqrt{\frac{ne^2}{2m^* \epsilon_0 \; \epsilon_{\mathrm{eff}(k)} k}},
\end{equation}
where $n$ is the electron density, $e$ is the elementary charge, $m^*$ is the effective electron mass in GaAs, $\epsilon_0$ is the vacuum permittivity, and $\epsilon_{\mathrm{eff}}(k)$ is the effective dielectric constant for wavenumber k \cite{Stern1967, Aizin2012, Yoon2014}. 
The corresponding plasmon velocity is given by $v_P = \omega_P/k$.
In the gated region, the metallic gate screens the electric field of the plasmon, resulting in a reduced propagation velocity. In this regime, the effective dielectric constant becomes:
\begin{equation}
    \epsilon_{eff}(k) \; = \; \frac{1}{2} \; [\epsilon_2 + \epsilon_1 \; \text{coth}(kd)] , 
\end{equation}
where $\epsilon_1 (\epsilon_2)$ is the dielectric constant of AlGaAs (GaAs), and $d$ is the distance between the 2DEG and the metallic gate \cite{Dyer2012}.\\

The white dashed line shown in Fig. \ref{gated_plasmon}c corresponds to the analytical calculation of the first plasmonic mode (N = 1), using \(d = \SI{100}{\nano\meter}\) and \(n = \SI{4.3e11}{\per\centi\meter\squared}\). This carrier density is approximately \SI{18}{\percent} higher than the nominal value provided by the wafer growth sheet (\SI{3.66e11}{\per\centi\meter\squared}).
This discrepancy may originate from the relatively high optical power used during the measurements, which may increase the carrier density, as well as from the finite frequency resolution of the FFT associated with the limited time window.
Together, the experimental observations and the analytical calculation demonstrate the excitation of plasmons in the 2DEG and their confinement within the gated cavity.

\newpage

\subsection{One-dimensional plasmonic excitation}

\begin{figure}[h]
    \centering
    \includegraphics[scale = 0.8]{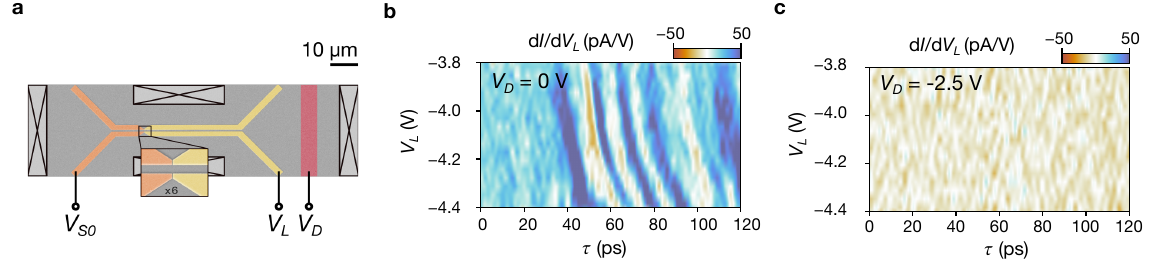}
    \caption{
    \textbf{1D THz plasmonic excitation measurement in the long electronic waveguide without and with the depletion gate.}
    \textbf{a.} Scanning electron microscope (SEM) image of the quantum device, showing the short (colored in orange) and long (colored in yellow) electronic waveguide, and the depletion gate colored un pink.
    \textbf{b.} Control measurement with the depletion gate set at 0V.
    Two-dimensional map of the time-resolved transmitted signal, derived with respect to the applied voltage $V_L$ on the long electronic waveguide.
    In this configuration we reproduce the data presented in Fig 3 in the main text.
    \textbf{c.} Reference measurement with the depletion gate set to -2.5 V, where the 2DEG is fully depleted and no current can flow. Two-dimensional map of the time-resolved transmitted signal as a function of the voltage $V_L$.
    }
    \label{1D_depletion} 
\end{figure}

\noindent To further confirm that the signals presented in Fig. 3 of the main text is the result of a 1D plasmonic excitation in the electronic waveguides and not possible parasitic contributions, we use the depletion gate to isolate the quantum device.
The first control experiment consists of repeating the conductance-dependent measurements of the long electronic waveguide , (L = \SI{30}{\micro\metre}), as presented in the main text, with and without depletion of the 2DEG beneath the depletion gate. 
The corresponding results are shown in Fig \ref{1D_depletion}, together with an SEM image of the device highlighting the long electronic waveguide and the depletion gate.
The time-resolved signal $dI/dV_L$ is first measured with the depletion gate grounded (Fig.\ref{1D_depletion}b), corresponding to the configuration used in the main text. 
The same measurement is then repeated with a depletion gate voltage of \SI{-2.5}{\volt} (Fig.\ref{1D_depletion}c), well beyond the pinch-off voltage.
In this fully depleted regime, the interference fringes completely disappear. 
This clearly demonstrates that the observed fringes originate from plasmonic excitations propagating through the one-dimensional electronic waveguide, and not from parasitic electromagnetic coupling.

\begin{figure}[h]
    \centering
    \includegraphics[scale = 1.0]{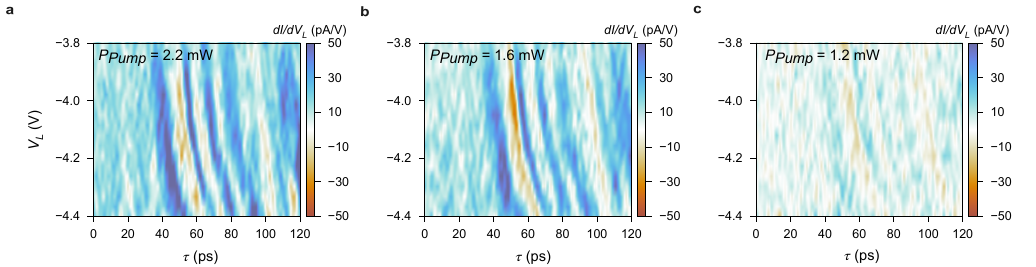}
    \caption{
    \textbf{Dependence of the pump laser power on the one-dimensional THz plasmonic excitation.}
    Two-dimensional maps of the time-resolved transmitted signal, differentiated with respect to the voltage $V_L$ applied to the long electronic waveguide, for different pump laser powers (indicated in red): \textbf{a} \(P_{\mathrm{pump}} = \SI{2.2}{\milli\watt}\), \textbf{b} \(P_{\mathrm{pump}} = \SI{1.6}{\milli\watt}\), and \textbf{c} \(P_{\mathrm{pump}} = \SI{1.2}{\milli\watt}\). 
    The optical power on the probe photoconductive switch is kept constant at \SI{1.7}{\milli\watt}, while PCS1 and PCS2 are biased at \SI{12}{\volt}.
    }
    \label{laser_dep} 
\end{figure}

A second control experiment is performed by repeating the same measurement while decreasing the optical power on the pump photoconductive switch. 
The measurements of $dI/dV_L$ are carried out for three different pump laser powers, $P_{\mathrm{pump}}$ = \SI{2.2}{\milli\watt}, \SI{1.6}{\milli\watt}, and \SI{1.2}{\milli\watt}, as shown in Fig. \ref{laser_dep}.
The amplitude of the signal decreases progressively as the pump laser power is reduced. 
This behaviour further confirms that the measured signal originates from the THz electrical pulse excitation of the 2DEG.

\clearpage
\newpage

\section{Extraction of the plasmon propagation velocity in the 1D electronic waveguide}

\begin{figure}[!h]
    \centering
    \includegraphics[scale = 0.8]{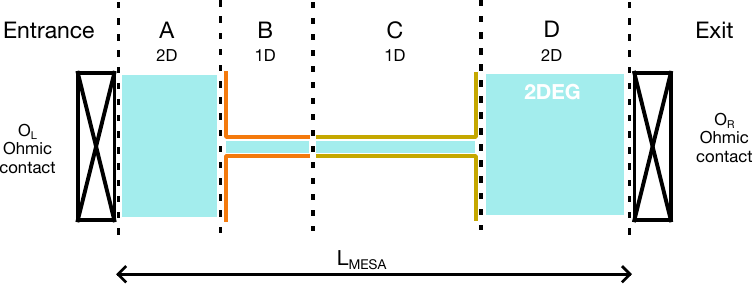}
    \caption{
    \textbf{Schematic and model for the quasi-1D plasmon velocity calculation.}
    Regions A and D correspond to the two-dimensional regions between the ohmic contacts $O_L$ and $O_R$.
    Regions B (C) correspond to the one-dimensional electronic waveguides with lengths of \SI{10}{\micro\meter} (\SI{30}{\micro\meter}).
    }
    \label{SM_velocity_schema}
\end{figure}

\noindent Using the propagation model introduced in the main text and illustrated by the schematic shown in Fig.\,\ref{SM_velocity_schema}, the velocity of the one-dimensional THz plasmonic excitation is extracted as a function of conductance.
The device is modelled as a sequence of four distinct regions. 
Regions A and D correspond to ungated regions of the two-dimensional electron gas (2DEG), while regions B and C correspond to the short and long electronic waveguides, respectively.
Within each region, we assume a uniform propagation velocity. 
The total propagation time is therefore expressed as the sum of four contributions, each associated with a well-defined propagation velocity.
In the two-dimensional map of the signal presented in Fig. 3 of the main text, the delay $\tau$ represents the time difference between the plasmonic excitation in the 2DEG and the electromagnetic signal travelling through the substrate.
More specifically, $t_{\mathrm{2DEG}}$ denotes the propagation time within the 2DEG, while $t_{\mathrm{substrate}}$ corresponds to the propagation time through the substrate beneath it,
\begin{equation}
    \tau = t_{\mathrm{2DEG}} - t_{\mathrm{substrate}}
\end{equation}
The propagation time $t_{\mathrm{2DEG}}$ is further expressed as the sum of contributions from the four regions defined in Fig. \ref{SM_velocity_schema}.
\begin{equation}
    t_{\mathrm{2DEG}}  \; = \; t_A \; + \; t_B \; + t_C \; + t_D \; = \; \frac{L_A}{v_A} \; + \; \frac{L_B}{v_B} \; + \;
    \frac{L_C}{v_C} \; + \;\frac{L_D}{v_D} \;
\end{equation}
with $L_{\rm{A-D}}$ and $v_{\rm{A-D}}$ being the corresponding lengths and velocities for each region.
The propagation velocity of the electromagnetic wave in the substrate is denoted $v_{\mathrm{bulk}}$.
The velocities in regions A and D are assumed to be the same and equal to the
previously reported 2D plasmon velocity $v_{\mathrm{p2D}}$ \cite{Wu2015, GeffroyPhD}.
By solving for the propagation time in each region, one can obtain two coupled equations for the velocities in region B (short waveguide) and region C (long waveguide), considering that the plasmon velocity depends on the conductance of the electronic waveguide \cite{Roussely2018, Takada2025},

\begin{equation}
    v_{C}(G_C)  \; = \; \frac{L_C}{\tau \; - \; \frac{(L_{A} + L_{D})}{v_{\mathrm{p2D}}} \; - \; \frac{L_B}{v_B(G_B)} 
    \; +\; \frac{L_{\mathrm{mesa}}}{v_{\mathrm{bulk}}} }
\end{equation}

\begin{equation}
    v_{B}(G_B)  \; = \; \frac{L_B}{\tau \; - \; \frac{(L_{A} + L_{D})}{v_{\mathrm{p2D}}} \; - \; \frac{L_C}{v_C(G_C)} 
    \; +\; \frac{L_{\mathrm{mesa}}}{v_{\mathrm{bulk}}} }
\end{equation}

Equations (5) and (6) allow us to determine the propagation velocity of the plasmonic excitation in the 1D electronic waveguide, denoted $v_{1D}$ (referred to as $v_{B}$  or $v_{C}$ depending on the waveguide segment in the device geometry).
These equations contain three unknown parameters: 
$v_{\mathrm{p2D}}$, $v_{B}(G)$ and $v_{C}(G)$. 
The propagation velocity in the two-dimensional electron gas, $v_{\mathrm{p2D}}$, has previously been measured to be $v_{\mathrm{p2D}}$ = $1.3 \times 10^7~\mathrm{ms^{-1}})$\cite{Wu2015} , leaving only the velocities in the two one-dimensional waveguide sections to be determined.

To extract these velocities, we introduce the parameter $\delta$, defined as the temporal separation between successive interference fringes. As discussed below, the observed fringes originate exclusively from reflections between the interface separating the short and long electronic waveguides and the left ohmic contact. Consequently, $\delta$ is given by

\begin{equation}
    \delta \; = \; 2 \; \left( \frac{L_A}{v_{\mathrm{p2D}}} \; + \; \frac{L_B}{v_B(G)} \right)
\end{equation}
where $L_A$ and $L_B$ are defined in Fig.\,\ref{SM_velocity_schema}.
To determine the plasmon propagation velocity over the full accessible conductance range in the one dimensional waveguide, we investigate two complementary measurement configurations.

In configuration (i), the conductance of the short waveguide is fixed to 11\,$G_0$, while the conductance of the long waveguide is varied between 2 $G_0$ and 5$G_0$.
In configuration (ii), the conductance of the long waveguide is fixed at 5$G_0$, while the conductance of the short waveguide is varied between 5$G_0$ and 11$G_0$.
Together, these two configurations provide access to $v_{1D}$ across the entire conductance range investigated.

\begin{figure}[!h]
    \centering
    \includegraphics[scale = 1.0]{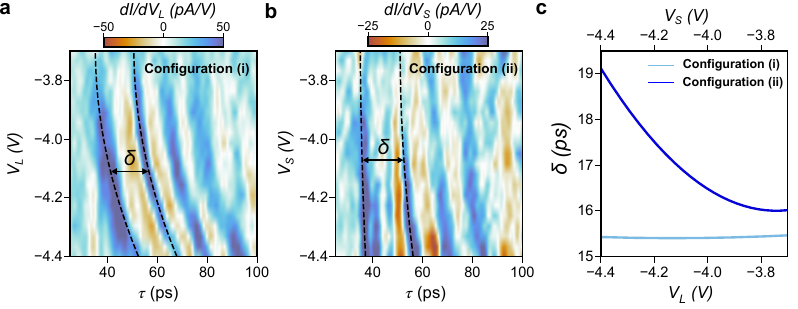}
    \caption{
    \textbf{Fringe separation used for the quasi-1D plasmon velocity calculation.}
    \textbf{(a,b)} Two-dimensional maps of the differential current as a function of delay time and gate voltage for configurations (i) and (ii), respectively.
    The arrows indicate the time separation $\delta$ between the two fringes.
    \textbf{(c)} Extracted fringe separation $\delta$ as a function of gate voltage for the two configurations.
    Configuration (i) shows an almost constant separation, while configuration (ii) shows a gate-dependent separation.
    }
    \label{SM_fringe_separation}
\end{figure}

The experimentally determined values of $\delta$ for both measurement configurations are shown in Fig. \ref{SM_fringe_separation}. 
In case (i), $\delta$ remains constant within the experimental uncertainty, whereas in case (ii), $\delta$ increases as the number of transmitting channels in the short waveguide is reduced. This behavior demonstrates that the interference fringes originate from reflections involving the short waveguide and the left ohmic contact, thereby ruling out reflections at interfaces C and D as the dominant contribution.

\begin{figure}[!h]
    \centering
    \includegraphics[scale = 0.8]{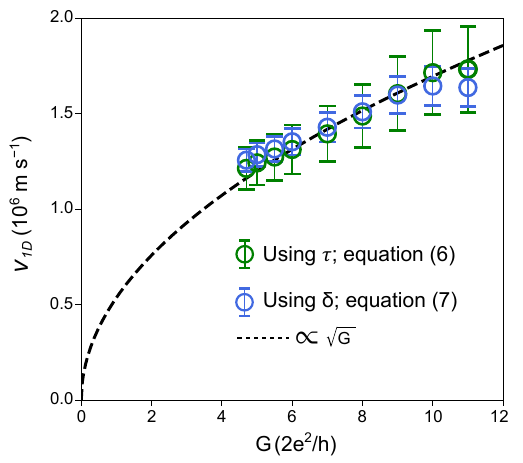}
    \caption{
    \textbf{Comparison of quasi-1D plasmon velocities extracted using two different methods.}
    The green circles show the velocity obtained using the fringe position $\tau$ in equation (6), while the blue squares show the velocity obtained using the fringe separation $\delta$ in equation (7).
    The black dashed line shows a square-root dependence, $\nu \propto \sqrt{G}$, used as a guide to the eye.
    The good agreement between the two methods confirms that the velocity extraction is consistent.
    }
    \label{SM_velocity_comparison}
\end{figure}

For configuration (i), the constant value of $\delta$ allows us to determine $v_B(11\,G_0)$ directly from Eq. (7). 
This value is then substituted into eq. (5), together with the experimentally measured delay times $\tau$, to extract $v_C(G)$ for conductances between $2G_0$ and $5G_0$.
For configuration (ii), we use equation (6) to determine the speed of the plamonic excitation $v_{1D}$.
Since $\delta$ now varies with the conductance of the short waveguide, the measured values of $\tau$, together with the previously determined value of $v_C(5\,G_0)$, allow us to determine $v_B(G)$ over the conductance range $(5G_0 \leq G \leq 11G_0)$. 
As an independent consistency check, the measured values of $\delta$ for configuration (ii) can also be used to calculate $v_B(G)$, yielding consistent results as shown in Fig. \ref{SM_velocity_comparison}.

The values extracted from both measurement configurations are combined in Fig. 4 of the main text to obtain the complete dependence of the one-dimensional plasmon propagation velocity $v_{1D}$ on conductance.%

In the table below we summarize the parameter which have been used in the velocity calculations.

\begin{table}
\end{table}
\begin{table}[h]
\centering
\begin{tabular}{c c c}
\hline
\textbf{Parameter} & \textbf{Value} & \textbf{Unit} \\
\hline
$L_A$ & 25 & $\mu$m \\
$L_B$ & 10 & $\mu$m \\
$L_C$ & 30 & $\mu$m \\
$L_D$ & 35 & $\mu$m \\
$L_{\mathrm{measa}}$ & 100 & $\mu$m \\
$v_{\mathrm{p2D}}$ & $1.3 \times 10^{7}$ & m/s \\

$v_{\mathrm{bulk}}$ & $9.1 \times 10^{7}$ & m/s \\
\hline
\end{tabular}
\caption{List of the paramters used in the velocity calculation.}
\label{tab:velocity_parameters}
\end{table}

\vspace{3mm}
\textbf{Dependence of velocity on channel number: }The velocity of the one-dimensional plasmon exhibits a square-root dependence on conductance and decreases as the conductance is reduced, in agreement with previous observations in both ungated and gated quasi-one-dimensional electronic waveguides \cite{Kukushkin2005, Roussely2018, Takada2025}.
This square-root dependence of the conductance can be understood using a transmission line model for 2D plasmons, which we extend to the quasi-one-dimensional regime for our case.
The effective width of our electronic waveguide is \SI{300}{\nano\meter} (see Method section), which is significantly larger than the Fermi wavelength of the 2DEG ($\lambda_F \approx \SI{13}{\nano\meter}$).
This allows multiple transverse electronic modes to remain populated within the electronic waveguide.
However, the plasmonic excitation remains effectively one-dimensional, since the wavelength of the plasmonic excitation in the \SIrange{0.1}{1}{\tera\hertz} range is on the order of \SI{1}{\micro\meter}, which is larger than the width of the electronic waveguide.

We consider the dispersion relation for gated plasmons and assume that it remains valid in the quasi-one-dimensional regime imposed by electrostatic confinement.
The plasmon velocity is then given by:

\begin{equation}
    v_P \; = \; \frac{\omega_P}{k} \; = \; \sqrt{\frac{ne^2}{m^* \epsilon_0 [\epsilon_2 + \epsilon_1 \text{coth}(kd)]k}} \; .
\end{equation}

\noindent From this expression, we introduce the capacitance per unit length, defined as $C = W \epsilon_0 [\epsilon_2 + \epsilon_1 \text{coth}(kd)]k$.
In the limit $kd \to 0$, this expression reduces to the standard gate capacitance of a HEMT, $C = W \epsilon_0 \epsilon_1 / d$ \cite{Dyer2012}.
Assuming that the sheet conductance is given by $G = ne\mu$, where $\mu$ is the mobility of the 2DEG, this relation remains valid in the quasi-one-dimensional limit.
The plasmon velocity can then be rewritten as: 

\begin{equation}
    v_P \; = \; \sqrt{\frac{ne^2W}{m^*C}} \; = \; \sqrt{\frac{eWG}{m^*C\mu}} \;.
\end{equation}

This square-root dependence is consistent with previous experimental observations in the GHz regime \cite{Kukushkin2005}, and demonstrates that the same electrostatic scaling persists for THz plasmonic excitations.

%

\end{document}